\documentclass[12pt]{article}
\pdfoutput=1
\usepackage{pstricks}
\usepackage{color}

\usepackage[english]{babel}
\usepackage{fancyhdr}
\usepackage{amsmath}
\usepackage{amssymb}
\usepackage{amsfonts}
\usepackage{psfrag}
\usepackage[applemac]{inputenc}
\usepackage{graphicx,wrapfig}
\usepackage[bf,footnotesize]{caption}
\usepackage{pstricks}
\usepackage{multicol}
\usepackage{enumerate}
\usepackage{color}
\usepackage{array}
\usepackage{url}
\usepackage[title,titletoc,toc]{appendix}
\usepackage{float}
 \usepackage{hyperref}
 \usepackage{graphicx}
 \usepackage{subfigure}
\usepackage{epstopdf}
\newcommand\Mark[1]{\textsuperscript#1}

\newenvironment{Eqnarray}{\arraycolsep 0.14em\begin{eqnarray}}{\end{eqnarray}}
\def\bea{\begin{Eqnarray}}
\def\eea{\end{Eqnarray}}

\newcommand{\beq}{\begin{equation}}
\newcommand{\eeq}{\end{equation}}
\newcommand{\bag}{\begin{align}}
\newcommand{\eag}{\end{align}}

\newcommand{\eV}{\,\mathrm{eV}}

\newcommand{\GeV}{\,\mathrm{GeV}}

\newcommand{\hc}{\,\mathrm{h.c.}}

\newcommand{\cw}{\cos\theta_W}
\newcommand{\sw}{\sin\theta_W}

\newcommand{\eq}[1]{Eq.~(\ref{#1})}
\newcommand{\fig}[1]{Fig.~\ref{#1}}
\newcommand{\eqs}[1]{Eqs.~(\ref{#1})}

\newcommand{\app}[1]{Appendix~\ref{#1}}

\newcommand{\nn}{\nonumber}

\newcommand{\HUone}{\hat H_{U_1}}
\newcommand{\HUtwo}{\hat H_{U_2}}
\newcommand{\HDone}{\hat H_{D_1}}
\newcommand{\HDtwo}{\hat H_{D_2}}
\newcommand{\sHUone}{H_{U_1}}
\newcommand{\sHUtwo}{H_{U_2}}

\newcommand{\fHUone}{\tilde H_{U_1}}
\newcommand{\fHUtwo}{\tilde H_{U_2}}

\newcommand{\diracmu}[1]{\mu^{\text{Dirac}}_{#1 \alpha}}
\newcommand{\diracmul}[1]{\mu^{\tilde \lambda}_{#1\alpha}}
\newcommand{\diracmub}[1]{\mu^{\tilde B}_{#1\alpha}}

\newcommand{\lambdak}{\tilde\lambda_k}

\begin{document}

\baselineskip=18pt

\setcounter{footnote}{0}
\setcounter{figure}{0}
\setcounter{table}{0}

%%%%%%%%%%%%%%%%%%%%%%%%%%%%%%%%%%%%%%%%%%%

%FRONTPAGE1%%%%%%
%\begin{flushright}
%UAB-FT--\\
%date
%\end{flushright}

%\title{Title}
%\author{Clara Peset}
%\affiliation{IFAE, Universitat Autonoma de Barcelona, 08193 Bellaterra, Barcelona}
%\begin{abstract}

%Abstract

%\end{abstract}
%\maketitle

%FRONTPAGE2%%%%%%
\begin{titlepage}
\begin{flushright}
%CU-LEPP--\\
%\today
\end{flushright}
\vspace{.3in}

\begin{center}
%\vspace{1cm}

{\Large \bf Neutrino masses in RPV models with \\[9pt] two pairs of Higgs doublets}

\vspace{.8cm}

%{\bf Yuval Grossman}\Mark{1}{\bf, Clara Peset}\Mark{2}

{\bf Yuval Grossman}\Mark{1} and {\bf Clara Peset$\,$}\Mark{2}

\vspace{.5cm}

\centerline{\Mark{1}{\it Laboratory for Elementary-Particle Physics, Cornell University, Ithaca, N.Y.}}

\vspace{.25cm}

\centerline{\Mark{2}{\it IFAE, Universitat Aut\`onoma de Barcelona, 08193 Bellaterra, Barcelona}}

\end{center}
\vspace{.8cm}

\begin{abstract}
\medskip
We study the generation of neutrino masses and mixing in supersymmetric
R-parity violating models containing two pairs of Higgs doublets. In these
models, new RPV terms $\HDone\HDtwo\hat E$ arise in the
superpotential, as well as new soft terms.
Such terms give new contributions to neutrino masses.
We identify the different parameters and suppression/enhancement
factors that control each of these contributions. At tree level, just
like in the MSSM, only one neutrino acquires a mass due to
neutrino-neutralino mixing. There are no new one loop effects. We study the 
two loop contributions and find the conditions under which they can be important.
\end{abstract}

\bigskip

\end{titlepage}

%%%%%%%%%%%%%%%%%%%%%%%%
\section{Introduction}

Neutrinos have a non-zero mass matrix, as is indicated by neutrino
oscillation experiments. This fact requires some extension of the
Standrad Model (SM) that incorporates both their masses and their
mixing angles~\cite{Neutrinos,Neutrinoexp,neutrinorev}. The experimental data~\cite{PDG},
\bea
&&~~~~~~\Delta m_{32}^2=\left(2.32^{+0.12}_{-0.08}\right) \times 10^{-3}\eV^2, \qquad \Delta m_{21}^2=(7.5\pm 0.20) \times
10^{-5}\eV^2, \\
&&\sin^2(2\theta_{32})>0.95, \qquad
\sin^2(2\theta_{12})=0.857\pm 0.024, \qquad \sin^2(2\theta_{13})=0.095\pm 0.010, \nonumber
\eea
exhibit a mild mass hierarchy, two large mixing angles, and one mixing angle
that is somewhat smaller.
This structure poses a challenge for new physics where, generally,
mass hierarchies come with small mixing angles. This is solved when
different neutrinos obtain their masses from different sources. Then,
cancellations in the determinant of the mass matrix can arise
naturally, making its value smaller than the typical values of the
elements of
the matrix. Neutrinos in R-Parity Violating (RPV) supersymmetric
models have been widely studied \cite{neutrinosRPV} and have been shown to
be a framework in which this property is accomplished. In these models one neutrino acquires a mass at tree level through neutrino-neutralino mixing, while the other two acquire their masses from loop effects. %\cite{Banks:1995by,Borzumati:1996hd,Grossman:2003gq}.
 
Models with extra Higgs doublets have been widely studied both in the
context of the SM~\cite{2HDM} and supersymmetry
(SUSY)~\cite{SUSY:2HDM}. In the SUSY case, the simplest way to ensure anomaly cancellation is to add pairs of down-type and
up-type Higgs fields. Lately, such models have been proposed as a way of
naturally lifting the mass of the lightest Higgs boson, which in the
Minimal Supersymmetric Model (MSSM) cannot be $125\GeV$ without some
amount of fine tuning~\cite{Alves:2012ez}. When R-parity is not
imposed in these models, new renormalizable terms of the form
$\hat H_{D_i}\hat H_{D_j}\hat E$ arise in the superpotential. Such new terms can
substantially contribute to the neutrino mass matrix since their couplings are less
constrained than the conventional leptonic RPV couplings.

In this work we study how neutrino masses arise in a general
supersymmetric model with more than the minimal number of Higgs doublets. The large number of free
parameters in the model does not allow to make predictions without any
kind of further assumption. Nevertheless, we identify the
suppression and enhancement factors in the various contributions
to the neutrino mass matrix. We find that, even with two pairs of
Higgs doublets, only one neutrino acquires a mass at tree level, just
like in the MSSM. 
We describe the loop diagrams generated by the new RPV terms in the
superpotential, which arise at the two loop level, and in the
appendix we give expressions for the relevant one loop diagrams within our
model. We study which of these diagrams may give relevant contributions to the neutrino masses.

One major issue in models with several Higgs doublets is that generally
they generate flavor changing neutral currents (FCNCs) that can cause
severe phenomenological problems. There are several ways to avoid such
bounds, for example by assuming a specific texture for the Yukawa
couplings to the quark sector, or by assuming Minimal Flavor
Violation (MFV), see, for example~\cite{2HDM}.  In this work we only
concentrate on the leptonic sector and thus we do not elaborate on the
quark sector, and just assume that one of the available solutions
to the FCNCs bounds is in place.

\section{The model}

We work with a general RPV low-energy supersymmetric model with one
extra pair of Higgs doublets, namely, we consider two up-type and two
down-type Higgs doublets. We follow the notation of
\cite{Grossman:1998py} where the model with just one pair of Higgs
doublets is fully described. Neutrino masses arise from diagrams which
violate lepton number by two units. In order to avoid the bounds from
proton stability, we choose only terms which still preserve ${\bf
Z_3}$ baryon triality~\cite{Ibanez:1991pr}. When R-parity is not imposed, the down-type Higgs supermultiplets $\HDone$ and $\HDtwo$ have the same quantum numbers as the lepton supermultiplets $\hat L_i$. We denote the five supermultiplets by one only symbol $\hat{L}_I$ ($I=0,1,2,3,4$) such that $\hat L_0\equiv \HDone$, $\hat L_1\equiv \HDtwo$ and $\hat L_{1+i}\equiv \hat L_i$. Throughout this work we will use the following index notation: upper-case Latin letters for the extended five-dimensional lepton flavor space, Greek letters for four-dimensional flavor spaces and lower-case Latin letters for three-dimensional ones.

The relevant renormalizable superpotential for this model is
\bea
W&=&\epsilon_{ij}\left[-\mu_{1I}\hat L_I^i \HUone^j-\mu_{2I}\hat L_I^i \HUtwo^j+\frac{1}{2}\lambda_{IJ m}\hat L_I^i \hat L_J^j \hat E_m+\lambda'_{I n m}\hat L_I^i \hat Q_n^j\hat D_m\right],\label{w}
\eea
where $\hat H_{U_i}\ , i=1, 2$, are the two up-type Higgs
supermultiplets, $\hat Q_n$ are the quark doublet supermultiplets,
$\hat U_m$ ($\hat D_m$) are the up-type (down-type) quark
supermultiplets, and $\hat E_m$ are the singlet charged lepton
supermultiplets. The $n$ and $m$ are
flavor indices. The coefficients $\lambda_{IJ m}$ are antisymmetric under the
exchange of the indices $I$ and $J$. The usual MSSM $\mu$-term is now
extended to two five-dimensional vectors, $\mu_{1I}$ and
$\mu_{2I}$. Note that, in comparison with the RPV models already
studied in \cite{Banks:1995by, Borzumati:1996hd, Grossman:2003gq}, a
new type of trilinear $\lambda$-term arises for the two down-type Higgs supermultiplets, which is less constrained than the conventional leptonic RPV terms,
\beq
\frac{\tilde\lambda_m}{2}\epsilon_{ij}\left(\HDone^i \HDtwo^j-\HDtwo^i \HDone^j\right)\hat E_m,\label{newterm}
\eeq
where $\tilde\lambda_m=\lambda_{01m}$.

In order to compute all the contributions to the neutrino masses we need to consider the following soft supersymmetry breaking terms:
\bea
V_{\mathrm{soft}}&=&\left(M_{\tilde L}^2\right)_{IJ} \tilde L^{i*}_I \tilde L^i_J-\left(\epsilon_{ij}B_{1I}\tilde L_I^i \sHUone^j+\hc \right)-\left(\epsilon_{ij}B_{2I}\tilde L_I^i \sHUtwo^j+\hc \right)\nn\\
&&+\epsilon_{ij}\left[\frac{1}{2}A_{IJ  m}\tilde L^{i}_I \tilde L^j_J \tilde E_m+A'_{I n m}\tilde L^{i}_I \tilde Q^j_n \tilde D_m+\hc\right],\label{V}
\eea
which correspond to the $A$-terms and $B$-terms of the superpotential
and the new scalar mass terms. The usual MSSM $B$-term is now extended
to a combination of five-dimensional vectors $B_{1I}$ and $B_{2I}$, and the MSSM single mass term for the down-type Higgs boson together with the $3 \times 3$ lepton mass matrix are now extended to a $5 \times 5$ matrix, $\left(M_{\tilde L}^2\right)_{IJ}$.
We also define
\beq
\langle H_{U_1}\rangle\equiv \frac{1}{\sqrt 2}v_{u_1}\ , \;\;\quad \quad \langle H_{U_2}\rangle\equiv \frac{1}{\sqrt 2}v_{u_2}\ , \;\;\quad \quad \langle \tilde\nu_I\rangle\equiv \frac{1}{\sqrt 2}v_I\ ,\;\; \quad \quad \label{vevs}
\eeq
\beq
v_u=\left(v_{u_1}^2+v_{u_2}^2\right)^{1/2}, \quad \quad v_d=\left(\sum v_I^2\right)^{1/2}, \quad \quad \mu_1=\left(\sum \mu_{1I}^2\right)^{1/2}, \quad \quad \mu_2=\left(\sum \mu_{2I}^2\right)^{1/2},\label{vud}
\eeq
with
\bea
v\equiv \left(\vert v_u\vert^2+\vert v_d\vert^2\right)^{1/2}=\frac{2 m_W}{g}=246 \GeV.\label{v}
\eea
The value of these vacuum expectation values can be determined by
minimizing the potential. Performing this minimization is beyond the
scope of this work.

%%%%%%%%%%%%%%%%%%%%%%%%%%%%%%%%%%%%%%%%%%%%%%%%%%%%%%%%%%%%%%%%%%%%%%%%%%%%%%%%%%%%
\section{Tree level neutrino masses}

The neutrino mass matrix receives contributions both from tree and
loop level effects. In this section we study the mass matrix that arises at tree level.

\begin{figure}[b]
\centering
\includegraphics[scale=0.6]{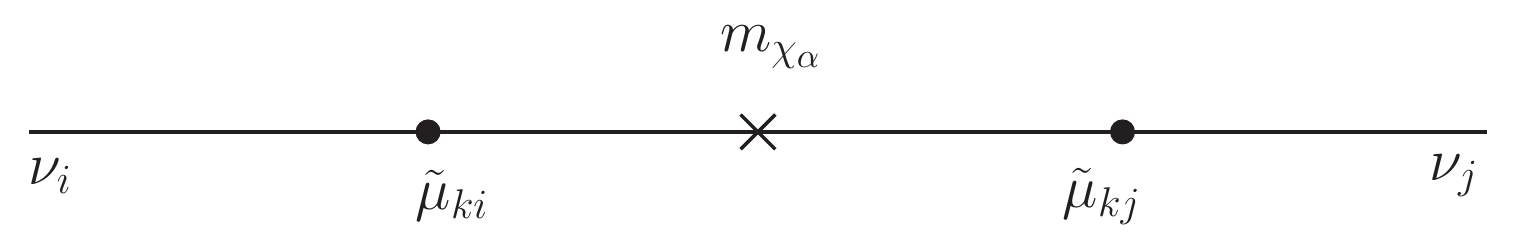}
\caption{Contribution to the tree level neutrino mass. The cross indicates a mass insertion for the neutralino with a Majorana mass. The blob indicates an RPV mixing.}
\label{tree level}
\end{figure}

The tree level masses arise from RPV mixing between the neutrinos and
the neutralinos, as shown in \fig{tree level}. Below we will first study
which are the alignment conditions of the five-dimensional expectation
value of $\hat L_I$ and the couplings $\mu_{1I}$ and $\mu_{2I}$, so
that the mass which arises at tree level is within the experimental
bounds. We will see how, even though we have doubled the number of Higgs-fields, still only one neutrino acquires a mass at tree level and we will give an explicit expression for that mass.

%\subsection{Alignment}
In our model we have a $9\times 9$ mass matrix for the neutralinos. In the basis $\{\tilde B, \tilde W, \fHUone,
\fHUtwo,\nu_I\}$, where we neglect the effects of non-renormalizable
operators, it is given by
\beq
M^N=
\begin{pmatrix} 
 M_1     & 0 & m_Z s_W \hat v_{u_1} & m_Z s_W \hat v_{u_2}& -m_Z s_W \hat v_I \\[3pt] 
 0 & M_2 & -m_Z c_W \hat v_{u_1} & -m_Z c_W \hat v_{u_2}& m_Z c_W \hat v_I\\[3pt]
m_Z s_W \hat v_{u_1} & -m_Z c_W \hat v_{u_1}& 0 & 0 & \mu_{1 I}  \\[3pt]
m_Z s_W \hat v_{u_2} & -m_Z c_W \hat v_{u_2}& 0 & 0 & \mu_{2I}  \\[3pt]
-m_Z s_W \hat v_I^T & m_Z c_W \hat v_I^T & \mu_{1I}^T &  \mu_{2I}^T & 0_{5\times5} \\
\end{pmatrix},\label{Mn}
\eeq
where $M_1$ is the Bino mass, $M_2$ is the Wino mass, $\hat v_x=v_x/v$, $c_W\equiv \cw$, $s_W\equiv \sw$ and $\theta_W$ is the Weinberg angle.
Note that none of the angles between the five-vectors $v_I$, $\mu_{1I}$ and $\mu_{2I}$ is small. The R-parity conservation limit corresponds to the case where the
three vectors are coplanar. Small R-parity breaking manifests itself
by the deviation of $v_I$ from the plain determined by $\mu_{1I}$ and $\mu_{2I}$.
Such deviation can be parametrized by the angle $\xi$ such that
\beq
\sin\xi=\frac{ (\hat \mu_{1} \times \hat \mu_{2}) \cdot \hat v}{\sin\chi},
\eeq
where $\hat a$ is a unit vector in the direction of the vector $a$ and
the angle 
\beq
\cos\chi=\frac{\sum \mu_{1I} \mu_{2I}}{\mu_1 \mu_2},
\eeq 
measures the alignment of $\mu_{1I}$ and $\mu_{2I}$. The cross product is defined on the three-dimensional space generated by the three five-vectors.
 
In order to find the masses, the first thing to note is that the mass matrix has rank seven and
thus there are two massless states at tree-level. The
product of the seven non-vanishing eigenvalues can be extracted from \eq{Mn}, and reads:
 \beq
 \mathrm{det'}M^N=2\frac{m_Z^2\, m_{\tilde\gamma}}{v^2} \,v_d^2
 \,\mu_1^2 \,\mu_2^2\sin^2\chi \sin^2\xi,
 \label{detm}\eeq
where we have defined $m_{\tilde \gamma}= M_1 c_W^2+M_2s_W^2$. Note that when  $v_I$, $\mu_{1I}$, and $\mu_{2I}$ are in the same
plane, $\xi=0$ and thus $\mathrm{det'}M^N=0$.

In order to get an estimate of the masses we consider the electroweak
breaking and SUSY breaking scales to be roughly equal and we denote them by
$\tilde m$. When we consider all the relevant masses to be of order
$\tilde m$ the product of the seven
non-vanishing masses should satisfy: $\mathrm{det'}M^N\leq \tilde m^6
m_3$. where $m_3$ is the mass of the heaviest neutrino.
In order for the neutrino masses to be within the experimental bounds
we thus require
\beq
\xi^2 \lesssim  \frac{m_3}{\tilde m},
\label{req1}
\eeq
where we used $\sin\chi \sim 1$.
We see that the expression we get is similar to the one for the case
of the MSSM \cite{Banks:1995by}. The small angle in the MSSM is the one
between the $\mu$ and $v$ vectors while here it is the angle between the plane
generated by the two $\mu$-like vectors and $v$.

In order to obtain the neutrino mass matrix we need to diagonalize the $9\times9$ matrix $M^N$. This computation is simplified by considering the hierarchical structure of the matrix to diagonalize:
\bea
M^N=\begin{pmatrix}
M_{6\times 6} & \mu_{6 \times 3}\\[4pt]
\mu^T_{3 \times 6} & 0_{3\times 3}
\end{pmatrix}\Longrightarrow
U M^N U^+=\begin{pmatrix}
M'_{6\times 6} & 0_{6 \times 3}\\[4pt]
0_{3 \times 6} & m_{\nu\;3\times 3}
\end{pmatrix},
\eea
where $M\gg\mu$ and therefore we may integrate out the six
neutralinos. 
From now on we work in the basis spanned by $\hat{L}_I$ such that $v_0=v_{d_1}$, $v_1=v_{d_2}$ and $v_m=0$ for $m=2,3,4$. Note that a basis in which all the $v_I$'s except one are zero could also be chosen, however, we prefer to keep our results in a more basis independent fashion.
To integrate out the neutralinos we use the see-saw mechanism, where $M$ is a Majorana mass and $\mu$ is a Dirac mass, and obtain the eigenvalues:
\beq
M'_{6\times 6}=M_{6\times 6},\qquad m_{\nu\;3\times 3}=\mu^T M^{-1} \mu.
\eeq
Now, defining the following ratios,
\beq
\frac{v_{d_1}}{v}=\cos \beta \cos\beta_1
,\qquad\frac{v_{d_2}}{v}=\cos\beta\sin\beta_1, \qquad
\frac{v_{u_1}}{v}=\sin\beta\cos \beta_2,\qquad\frac{v_{u_2}}{v}=\sin\beta\sin \beta_2,
\eeq
where $\beta$ is the usual angle defined by the ratio $v_u/v_d=\tan\beta$.
We find the neutrino mass matrix:
\bea
(m_{\nu})_{ij}&=&\frac{X}{\Delta \mu ^2}\left[\mu_{1i} \tilde\mu_2-\mu_{2i} \tilde\mu_1\right]\left[\mu_{1j}\tilde\mu_2-\mu_{2j}\tilde\mu_1\right]\label{numatrix}
\label{TLmass}
\eea
where
\beq
X\equiv\frac{m_{\tilde\gamma} m_Z^2 \cos ^2\beta}{  M_1 M_2\Delta \mu ^2+m_{\tilde\gamma} m_Z^2 \sin (2 \beta ) (\tilde\mu_1\sin \beta_2-\tilde\mu_2\cos \beta_2)}\sim \frac{\cos^2\beta}{\tilde m},
\label{X}
\eeq
and we have defined,
\beq
\tilde\mu_i\equiv\mu_{1d_i} \sin \beta_1-\mu_{2d_i}\cos \beta_1, \qquad
\Delta\mu^2\equiv\mu_{2d_2}\mu_{1d_1}-\mu_{1d_2}\mu_{2d_1}.
\eeq
In the last step of \eq{X} we have taken all the relevant masses to be
$\tilde m$. 

The tree level neutrino masses are the eigenvalues of the rank one matrix in \eq{numatrix} and therefore there is just one massive neutrino:
\beq
m_3= \frac{X}{\Delta\mu^2} (\tilde \mu_2 \vec \mu_1-\tilde \mu_1 \vec \mu_2)^2=\frac{X}{\Delta\mu^2} \mu_1^2 \mu_2^2 \sin^2\chi \sin^2 \xi,\qquad
m_1=m_2=0,
\eeq
where $\vec \mu_i=\mu_{ij}$. We define in the following $m_3>m_2>m_1$. As expected, the tree level neutrino mass is quadratically proportional to the small parameter that measures the RPV.

%%%%%%%%%%%%%%%%%%%%%%%%%%%%%%%%%%%%%%%%%%%%%%%%%%%%%%%%%%%%%%%%%
\section{Loop contributions to the neutrino mass matrix}

The neutrino mass matrix receives contributions from loop diagrams
with $\Delta L=2$. There are one loop contributions due to RPV
couplings that are present also in models with one pair of Higgs
doublets. They have already been thoroughly studied (see for example
\cite{Grossman:2003gq,Davidson1,Davidson2}), and we collect them in \app{oneloop} for
completeness.

Here we concentrate on the new diagrams that arise only once the second pair of
Higgs doublets is introduced. Strictly speaking, the only new term that is introduced is $\tilde
\lambda$. Yet, below we also consider effects that are due to the
extended $B$-term, namely $\tilde B$, which has been defined in \eqref{BiH}. 
We find that the new effects that are
generated by the new $\tilde \lambda$ term in the superpotential enter
the neutrino masses only at two loops. Roughly speaking, this is because the $ \tilde{\lambda} $ term does not involve any neutrinos. Thus it only breaks lepton number by one
unit in the
charged lepton sector and the transformation of this breaking into the
neutrinos appears at one loop. Since we need two of them, we end up with
a two loop effect.

The effects of the $\tilde B$ coupling on the neutrino mass matrix
arise both at one and two loops. The one loop effect is collected in
\app{oneloop}. Here we include some of the results for two loop diagrams in order to give an estimate of their possible importance.
In general, we expect such two loop effects to be smaller than the
one loop effects that the MSSM also presents. Yet,  
since the coefficients $\lambdak$ are less constrained than the usual RPV coefficients, these two loop diagrams could give important contributions to the neutrino mass matrix.

There are two types of effects that we call separable and
non-separable two loop contributions to the neutrino matrix. We
study them both below. 

\subsection{Separable contributions}
\begin{figure}[t]
     \begin{center}
        \subfigure[Effective coupling $\diracmul{i}$.]{%
            \label{Smulambda}
            \includegraphics[width=0.47\textwidth]{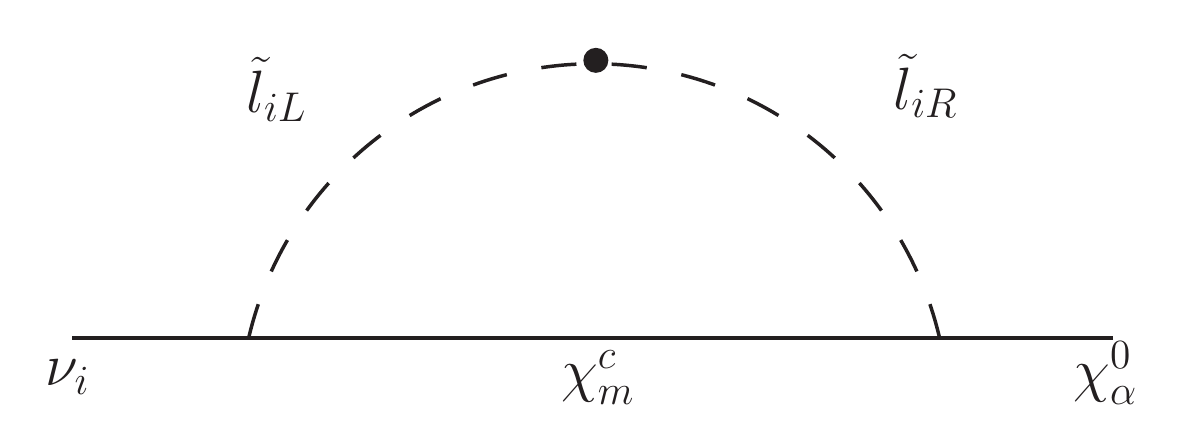} }
        \subfigure[Effective coupling $\diracmub{i}$. ]{%
           \label{SmuB}
           \includegraphics[width=0.47\textwidth]{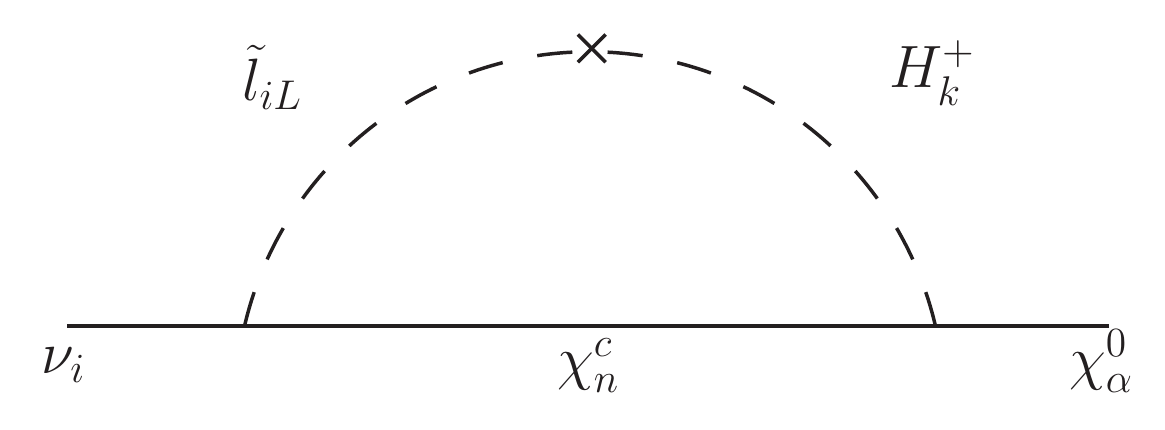} }
    \end{center}
    \caption{%
        The blob indicates the mixing between left and right-handed sleptons. The cross indicates the RPV B-vertex.
     }%
   \label{diracmu}
\end{figure}

For the separable contributions we study the Dirac-like
neutrino-neutralino mixing (see \fig{diracmu}). We define an effective coupling for this mixing at first order,
\beq
i\diracmu{i}=i\diracmul{i}+i\diracmub{i}.
\eeq
The effective coupling $\diracmul{i}$ corresponds to the diagram in \fig{Smulambda}, and can be expressed as:
\bea
\diracmul{i}
=\frac{ 1}{8\pi^2}\sum_m g\tilde\lambda_i(Z_-^{3m}Z_N^{4\alpha}-Z_-^{2m}Z_N^{5\alpha})^* Z_-^{1m} \frac{m_{\chi_m}\Delta^2 m_{\tilde l_i}}{ m_{\tilde l_i}^2}\approx \frac{3}{8\pi^2}g\tilde\lambda_im_{l_i},
\eea
where the $Z$'s refer to the appropriate mixing matrices defined as in the MSSM \cite{Dreiner:2008tw,Rosiek:1995kg,Martin:1997ns} but enlarged so that they accommodate the extra particle states of our model. In the last step, we have set $\Delta m_{\tilde l_i}^2\approx 2
m_{l_i} \tilde m$ and $m_{\tilde l_i}\sim m_{\chi_m}\sim \tilde m$.
The effective coupling $\diracmub{i}$ is represented in \fig{SmuB} and can be expressed as
\beq
\diracmub{i}=i \sum_{n,k} g^2 \tilde B_{ik} C_1^{\alpha nki}
m_{\chi_n}I_3(m_{\chi_n},m_{\tilde l_i},m_{H_k})\approx
\sum_{k}\frac{3}{64\pi^2}g^2 \frac{\tilde B_{ik}}{\tilde m} ,
\eeq
where
\beq
C_1^{\alpha nki}\equiv \tilde
Z_-^{1n}(Z_H^{2k}+Z_H^{3k})\left[(Z_N^{4\alpha}+Z_N^{5\alpha})Z_-^{1i}-\frac{1}{\sqrt
    2}\left(Z_N^{1\alpha}+\frac{g'}{g}Z_N^{0\alpha}\right)(Z_-^{2i}+Z_-^{3i})\right],
\eeq
$\tilde B_{ik}$ is defined in \eq{Bik} and $I_3$ 
is given in \eq{I3} (\eq{I3m} for the equal masses case). In the final step we have taken all the masses to
be at the supersymmetry breaking scale and we use $C_1^{\alpha nki}\sim 0.5$.

%\subsection{Sparable two loop contributions}

%\begin{figure}[H]
%\centering
%\includegraphics[scale=.55]{Diagrams/2Hlambdalambda.pdf}
%\caption{Loop from trilinear vertex with 2 Higses:$\lambda\lambda$ type}
%\label{2Hlambdalambda}
%\end{figure}

The separable contribution to the neutrino mass matrix that is proportional to the coupling $\tilde\lambda\tilde\lambda$ is
\beq \label{eq:llest}
\left[m_\nu\right]_{ij}^{\text{S},\tilde\lambda\tilde\lambda}=\sum_\alpha\frac{\diracmul{i}\diracmul{j}}{m_{\chi_\alpha^0}}\approx \frac{27}{32\pi^4}g^2\tilde\lambda_i\tilde\lambda_j\frac{m_{l_i}m_{l_j}}{\tilde m},
\eeq
where we used the approximation $m_{\chi_\alpha}\sim \tilde m$. This contribution is suppressed by two loop factors, two
RPV couplings and two leptonic Yukawa couplings. The latter makes this
contribution irrelevant in most cases.

Moving to the one that depends on  $\lambdak\tilde B$ we get
\beq \label{eq:lBest}
\left[m_\nu\right]_{ij}^{\text{S},\tilde\lambda\tilde B}=\sum_\alpha\frac{\diracmul{i}\diracmub{j}+\diracmub{i}\diracmul{j}}{m_{\chi_\alpha^0}}\approx \sum_k\frac{27}{128\pi^4}g^3 \frac{\tilde\lambda_i \tilde B_{jk}m_{l_i}+\tilde\lambda_j \tilde B_{ik}m_{l_j}}{\tilde m^2},
\eeq
where in the last step we consider $m_{\chi_\alpha}\sim \tilde m$. The suppression factors in this case are given by two loop factors, one
Yukawa coupling and the two RPV couplings $\tilde\lambda$ and $\tilde B$. 

Last we show the result for the loop that depends on $\tilde B \tilde
B$. It is given by 
\beq \label{eq:BBest}
\left[m_\nu\right]_{ij}^{\text{S},\tilde B \tilde B}=\sum_\alpha\frac{\diracmub{i}\diracmub{j}}{m_{\chi_\alpha^0}}\approx \sum_{k, k'}\frac{27}{512\pi^4}g^4 \frac{\tilde B_{ik}\tilde B_{jk'}}{\tilde m^3},
\eeq
where in the last step $m_{\chi_\alpha}\approx \tilde m$ is considered. The suppression factors in this case are given by two loop factors and the two RPV couplings $\tilde\lambda$, $\tilde B$. Since there is no leptonic Yukawa coupling in this case, this is the least suppressed of these contributions.
\begin{figure}[t]
     \begin{center}
        \subfigure[$\tilde B\tilde \lambda$ diagram]{%
            \label{NSBlambda}
            \includegraphics[width=0.45\textwidth]{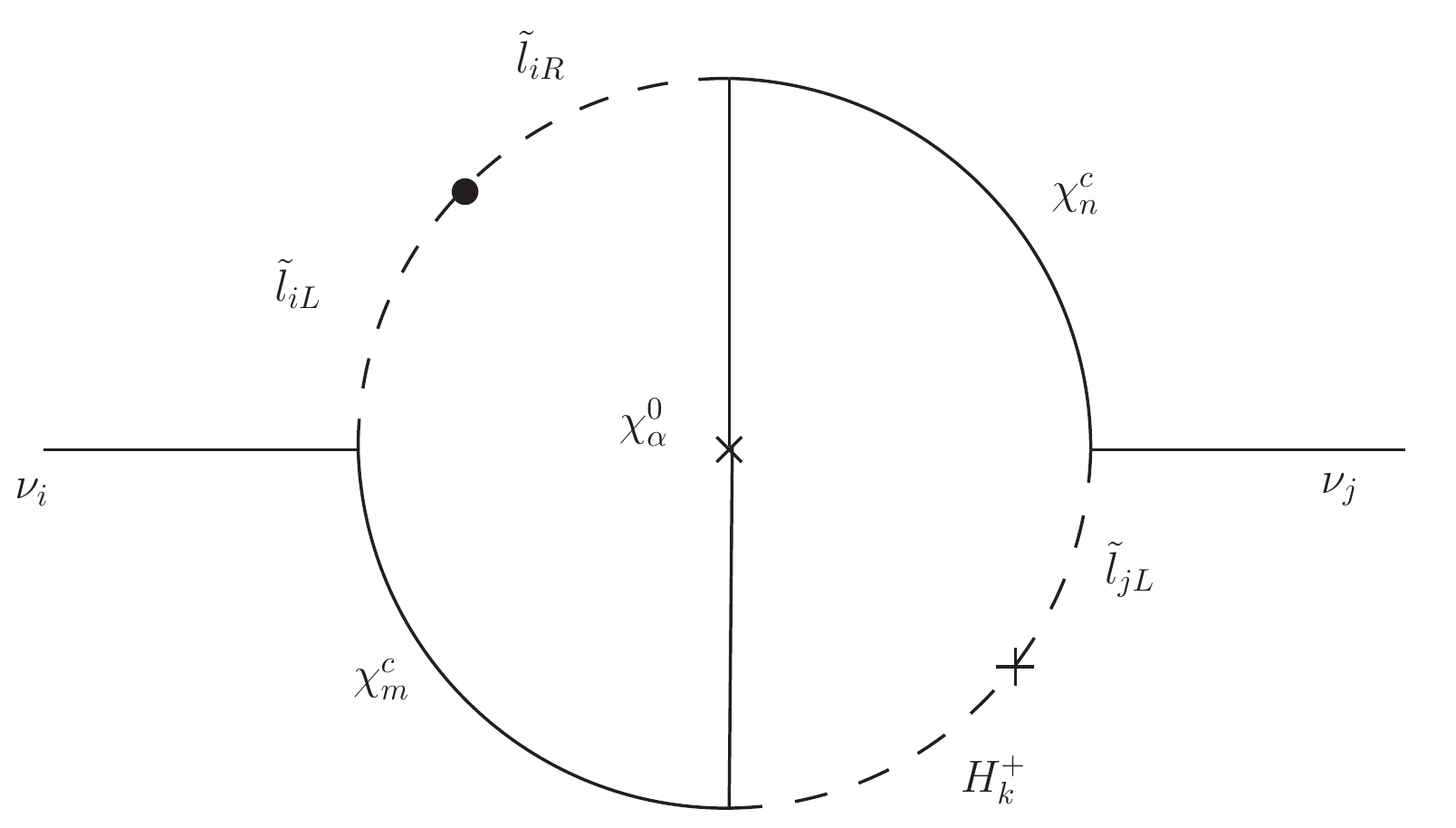} }
        \subfigure[$\tilde \lambda\tilde \lambda$ diagram]{%
           \label{NSll}
           \includegraphics[width=0.45\textwidth]{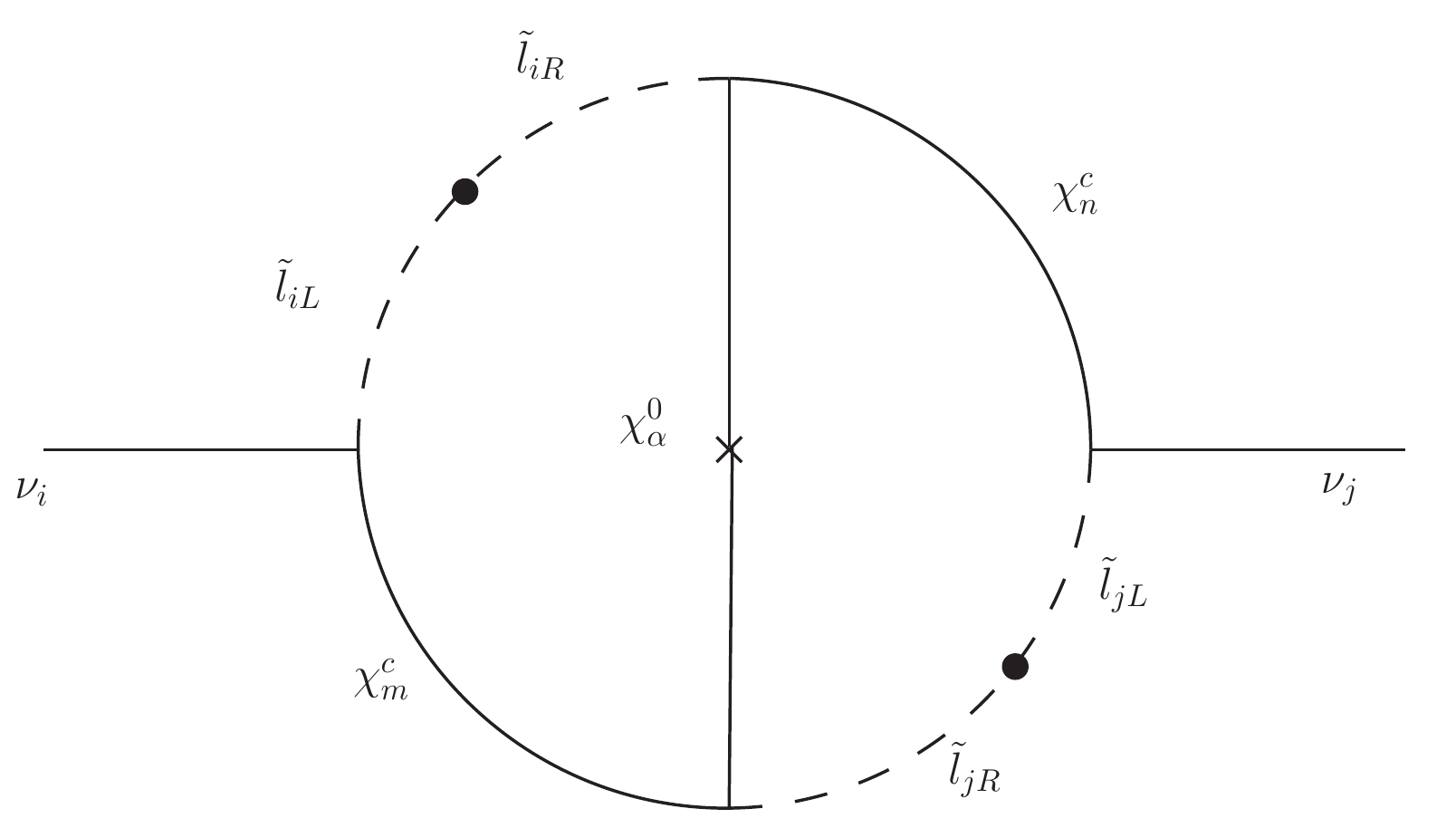} }
           \subfigure[$\tilde B\tilde B$ diagram]{%
           \label{NSBB}
           \includegraphics[width=0.45\textwidth]{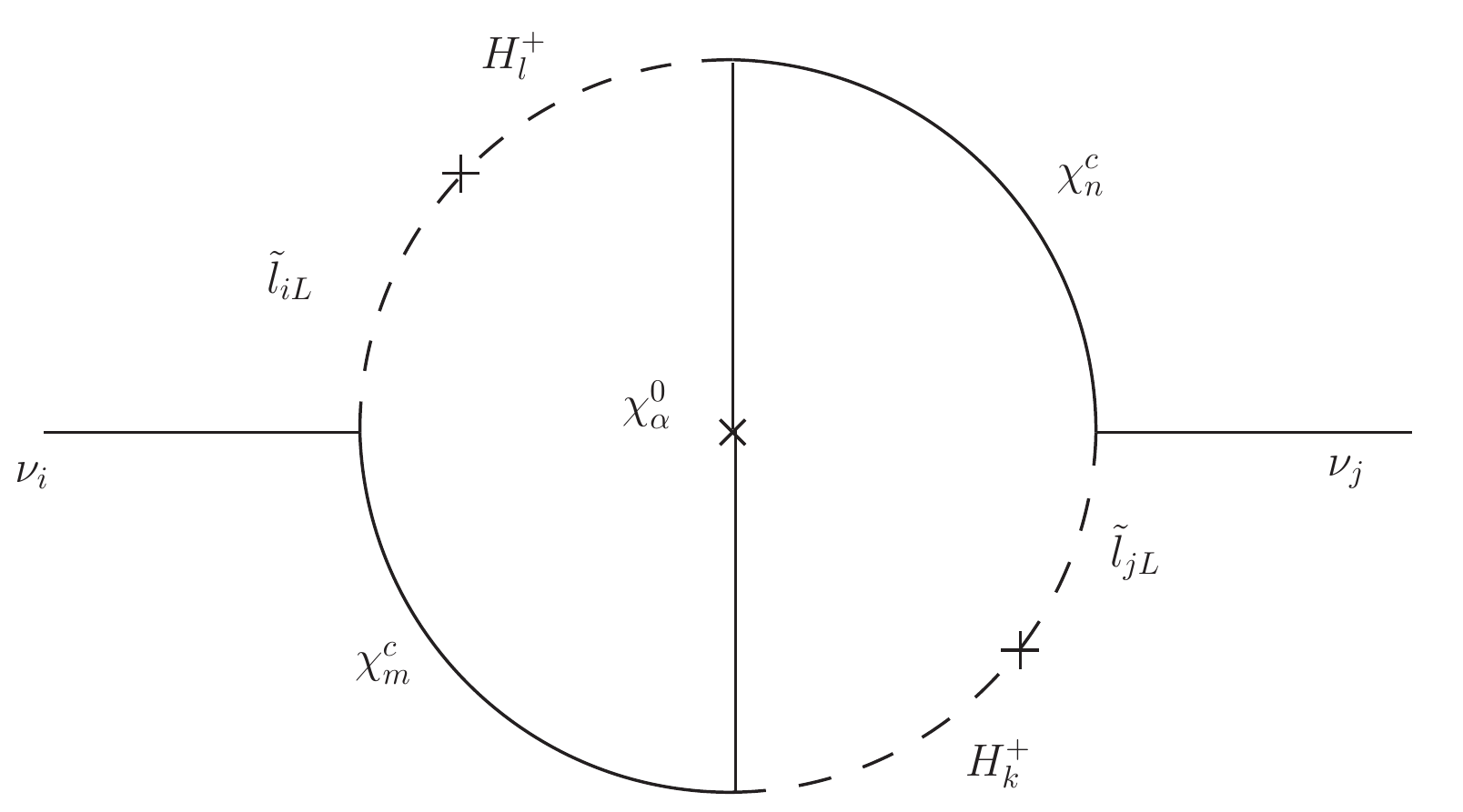} }
    \end{center}
    \caption{ Non-separable two loop diagrams that contribute to neutrino
  masses. The cross in the bosonic line indicates the RPV $B$ vertex.}
   \label{2lns}
\end{figure}

\subsection{Non-separable contributions}

We now move to discuss non-separable two loop diagrams. We have found that there are several of them. We include here three representative cases in order to have an insight of their possible importance.
These diagrams are represented in \fig{2lns}, and we discuss them in
turn below.

For the $\tilde B\tilde \lambda$-diagram in \fig{NSBlambda} we find
\beq
\left[m_\nu\right]_{ij}^{\text{NS},\tilde\lambda\tilde B}=
\sum_{\alpha,n,m,k}2g^3\tilde\lambda_i^*\tilde B_{jk}m_{\chi_\alpha}m_{\chi_m}m_{\chi_n}\Delta 
m_{\tilde l_i}^2 C_2^{\alpha m n k i} I_6(m_{\tilde l_i},m_{\chi_m},m_{\chi_n},m_{\tilde l_j},m_{H_k},m_{\chi_\alpha}),
\eeq
where
\bea
&&\!\!\!\! C_2^{\alpha m n k i} \equiv \\
&&\!\!\!\! Z_-^{1m}Z_-^{1n}(Z_H^{2k}+Z_H^{3k})^*(Z_-^{3n}Z_N^{4\alpha}-Z_-^{2n}Z_N^{5\alpha})^*
\left[(Z_N^{4\alpha}+Z_N^{5\alpha})Z_-^{1i}-\frac{1}{\sqrt
    2}\left(Z_N^{1\alpha}+\frac{g'}{g}Z_N^{0\alpha}\right)(Z_-^{2i}+Z_-^{3i})\right]
\nonumber
\eea
and $I_6$ is defined in \eq{eq:I6}. Note that $C_2^{\alpha m n k i}$ has several subtractions of $Z$'s and so it could undergo large cancellations.
Taking all the masses to be at the electroweak scale, and using
$C_2^{\alpha m n k i}\sim 0.5$, we find
\beq \label{eq:non-est}
\left[m_\nu\right]_{ij}^{\text{NS},\tilde\lambda\tilde B}\approx 
-\sum_{k}\frac{15.12}{256\pi^4} g^3\tilde\lambda_i^*\tilde B_{jk} 
\frac{m_{l_i}}{\tilde m^2},
\eeq
where $I_6$ for the equal masses case has been computed in \eq{I6m}.
This contribution to the neutrino mass matrix is suppressed by a
lepton mass, the trilinear RPV $\tilde\lambda$-coupling, the bilinear
supersymmetry-breaking RPV $\tilde B$-coupling, and two loop factors.

Moving to the $\tilde \lambda\tilde \lambda$-diagram in \fig{NSll} we obtain
\beq
\left[m_\nu\right]_{ij}^{\text{NS},\tilde\lambda\tilde \lambda}=
-\sum_{\alpha,n,m,k}4 g^2\tilde\lambda_i^*\tilde \lambda_j^* m_{\chi_\alpha}m_{\chi_m}m_{\chi_n}\Delta 
m_{\tilde l_i}^2\Delta 
m_{\tilde l_j}^2 C_3^{\alpha m n} I_5(m_{\tilde l_i},m_{\chi_m},m_{\tilde l_j},m_{\chi_n},m_{\chi_\alpha}),
\eeq
where,
\beq
C_3^{\alpha m n} \equiv Z_-^{1m}Z_-^{1n}(Z_-^{3n}Z_N^{4\alpha}-Z_-^{2n}Z_N^{5\alpha})^*(Z_-^{3m}Z_N^{4\alpha}-Z_-^{2m}Z_N^{5\alpha})^*
\eeq
and $I_5$ is defined in \eq{eq:I5}. Taking all the masses to be at the electroweak scale, and considering $C_3^{\alpha m n }\sim 0.5$, we find:
\beq \label{eq:non-est2}
\left[m_\nu\right]_{ij}^{\text{NS},\tilde\lambda\tilde \lambda}\approx 
\frac{60.48}{256\pi^4} g^2\tilde\lambda_i^*\tilde\lambda_j^*
\frac{m_{l_i}m_{l_j}}{\tilde m},
\eeq
where $I_5$ for the equal mass case has been computed in \eq{I5m}.
This contribution to the neutrino mass matrix is suppressed by two
lepton masses, two trilinear RPV $\tilde\lambda$-couplings, and two loop factors.

Finally, for the $\tilde B\tilde B$-diagram in \fig{NSBB}, the result reads
\beq
\left[m_\nu\right]_{ij}^{\text{NS},\tilde B\tilde B}=-
\sum_{\alpha,n,m,k,l}g^2\tilde B_{il}\tilde B_{jk}m_{\chi_\alpha}m_{\chi_m}m_{\chi_n} C_4^{\alpha m n k i} I_7(m_{\chi_m},m_{H_l},m_{\tilde l_i},m_{\chi_n},m_{H_k},m_{\tilde l_j},m_{\chi_\alpha}),
\eeq
where,
\bea
C_4^{\alpha m n k i j} &\equiv& Z_-^{1m}Z_-^{1n}(Z_H^{2k}+Z_H^{3k})^*(Z_-^{3n}Z_N^{4\alpha}-Z_-^{2n}Z_N^{5\alpha})^*\\
&&\left[(Z_N^{4\alpha}+Z_N^{5\alpha})Z_-^{1i}-\frac{1}{\sqrt
    2}\left(Z_N^{1\alpha}+\frac{g'}{g}Z_N^{0\alpha}\right)(Z_-^{2i}+Z_-^{3i})\right]\nn\\
    &&(Z_H^{2l}+Z_H^{3l})^*(Z_-^{3m}Z_N^{4\alpha}-Z_-^{2m}Z_N^{5\alpha})^*
\left[(Z_N^{4\alpha}+Z_N^{5\alpha})Z_-^{1j}-\frac{1}{\sqrt
    2}\left(Z_N^{1\alpha}+\frac{g'}{g}Z_N^{0\alpha}\right)(Z_-^{2j}+Z_-^{3j})\right] \nn
\eea
and $I_7$ is defined in \eq{eq:I7}. Note that $C_4^{\alpha m n k i j}$, just as $C_2^{\alpha m n k i}$, has several subtractions of $Z$'s and so it could also undergo large cancellations.
Taking all the masses to be at the electroweak scale, and using
$C_4^{\alpha m n k i j}\sim 0.5$, we find: 
\beq \label{eq:non-est3}
\left[m_\nu\right]_{ij}^{\text{NS},\tilde B\tilde B}\approx 
\sum_{k,l}\frac{3.80}{256\pi^4} g^2\frac{\tilde B_{il}\tilde B_{jk}}{\tilde m^3},
\eeq
where $I_7$ for the equal masses case has been computed in \eq{I7m}.
This contribution to the neutrino mass matrix is suppressed by two bilinear
supersymmetry-breaking RPV $\tilde B$-couplings, and two loop factors. Note that there is no Yukawa suppression for this diagram.

%%%%%%%%%%%%%%%%%%%%%%%%%%%%%%%%%%%%%%%%%%%%%%%%%%%%%%%%%%%%%%%%%%%
\section{Conclusions}

We study new sources of neutrino masses in RPV supersymmetric models with an extra pair of Higgs
doublets. In these models there is a new type of RPV term in the
superpotential of the form $\tilde \lambda_{k} \hat H_{D_1}\hat H_{D_2}E_k$. Such
a term is forbidden
in the MSSM since $ \lambda$ is antisymmetric in its first two
indices. There
are also similar new soft SUSY breaking terms.  
These new terms violate lepton number by one unit and therefore two such
terms can induce Majorana neutrino masses.

We find that the tree level effects that arise due to neutrino-neutralino mixing, contribute to the mass of only one neutrino, just like it happens in the MSSM. The value of this mass is quadratically proportional to the small R-parity breaking parameter, which in this case is measured by the deviation of the vector $v$ from planarity with respect to the two $\mu$-like vectors.

At the loop level we find that the new term can contribute to the mass matrix only through
two loop diagrams. Thus, in general we expect such terms not to be
significant. The estimates of the different diagrams are given in
Eqs. (\ref{eq:llest}), (\ref{eq:lBest}), (\ref{eq:BBest}),
(\ref{eq:non-est}), (\ref{eq:non-est2}), and (\ref{eq:non-est3}). Since they depend on different RPV parameters it is
not always clear which one gives the most important contribution. There is,
however, one factor that tells them apart which is the amount of
Yukawa suppressions. We see that the number of Yukawa factors is the
same as the number of $\tilde \lambda$ couplings. 

If we make the
assumption that all RPV parameters are of the same order,
that is, $\tilde B/\tilde m^2 \sim \tilde \mu/\tilde m \sim \tilde \lambda$, the Yukawa
suppression governs the hierarchy. In that case the diagrams without
any $\tilde\lambda$ couplings are the most important, that is, \eqs{eq:BBest} and (\ref{eq:non-est3}) are expected
to give the dominant effect. Nevertheless, the are one loop effects proportional to two $\tilde B$'s as in \eq{1LBB} and thus it is unlikely that the
two loop effects will be important. 

On the other hand, if we consider another plausible
assumption, namely that the only coupling that
is significant is $\tilde\lambda$, we find that its effect is always suppressed
by one small Yukawa, and so it can be important only when
$\tilde \lambda$ is very large. In this case, we could consider $\tilde B/\tilde m \sim \tilde \mu\sim \tilde \lambda m_l$ and so the leading contributions will be Eqs. (\ref{eq:llest}), (\ref{eq:lBest}),
(\ref{eq:non-est}), and (\ref{eq:non-est2}).

Our results can be extended to other similar models. They include
models where the extra Higgs states are not just simple duplication of
the MSSM one. They may be relevant also to a case study
in~\cite{Csaki:2013jza} where non-holomorphic terms like
$EH_DH_U^\dagger$ can appear. 

To conclude, neutrino masses can be used to put bounds on any model
with lepton number violation. In the model we considered, due to the fact that the new term we
study couples only to right handed charged leptons, its contribution to
neutrino masses is somewhat suppressed.  Thus, neutrino masses may not
give severe bounds on such models.

%%%%%%%%%%%%%%
\section*{Acknowledgments}
We thank Daniele Alves, Jeff Dror, Javi Serra, and Tomer Volansky for helpful discussions.
YG is a Weston Visiting Professor at the Weizmann Institute.
This work was partially supported by a grant from the Simons
Foundation ($\#$267432 to Yuval Grossman).
The work of YG is supported is part by the U.S. National
Science Foundation through grant PHY-0757868 and
by the United States-Israel Binational Science
Foundation (BSF) under grant No.~2010221. CP thanks Cornell University for hospitality during the course of this work. The work of CP was partially supported by the Universitat Aut\`onoma de Barcelona PR-404-01-2/E2010.

%%%%%%%%%%%%%%%%%%%%%%%%%%%%%%%%%%%%%%%%%%%%%%%%%%%%%%%%%%%%%%%%%%%%%%%%%%%%%%%%%%%%%55
\newpage
\begin{appendices}

\section{Feynman Rules}
\label{Feynrules}
In this Appendix we give the set of Feynman rules in our model necessary for describing all the diagrams studied in this work. As a reference for notation we have followed the MSSM Feynman rules in \cite{Dreiner:2008tw}. For every rule described here, there is one with all arrows reversed and complex conjugated couplings (except for the explicit $i$).
In all the cases, fermions are taken to be in their eigenstate basis and sfermions in a basis where they are their supersymmetric partners.

In this model there are RPV bilinear $\mu$-like terms involving a
neutrino which arise from \eq{w}, RPV bilinear terms involving neutral
scalars and RPV bilinear terms involving charged scalars, both arising
from \eq{V}. These vertices and their Feynman rules are represented
below
\vspace*{3mm}

\begin{minipage}{0.45\linewidth}   
  %  \begin{figure}[H]
\includegraphics[scale=.45]{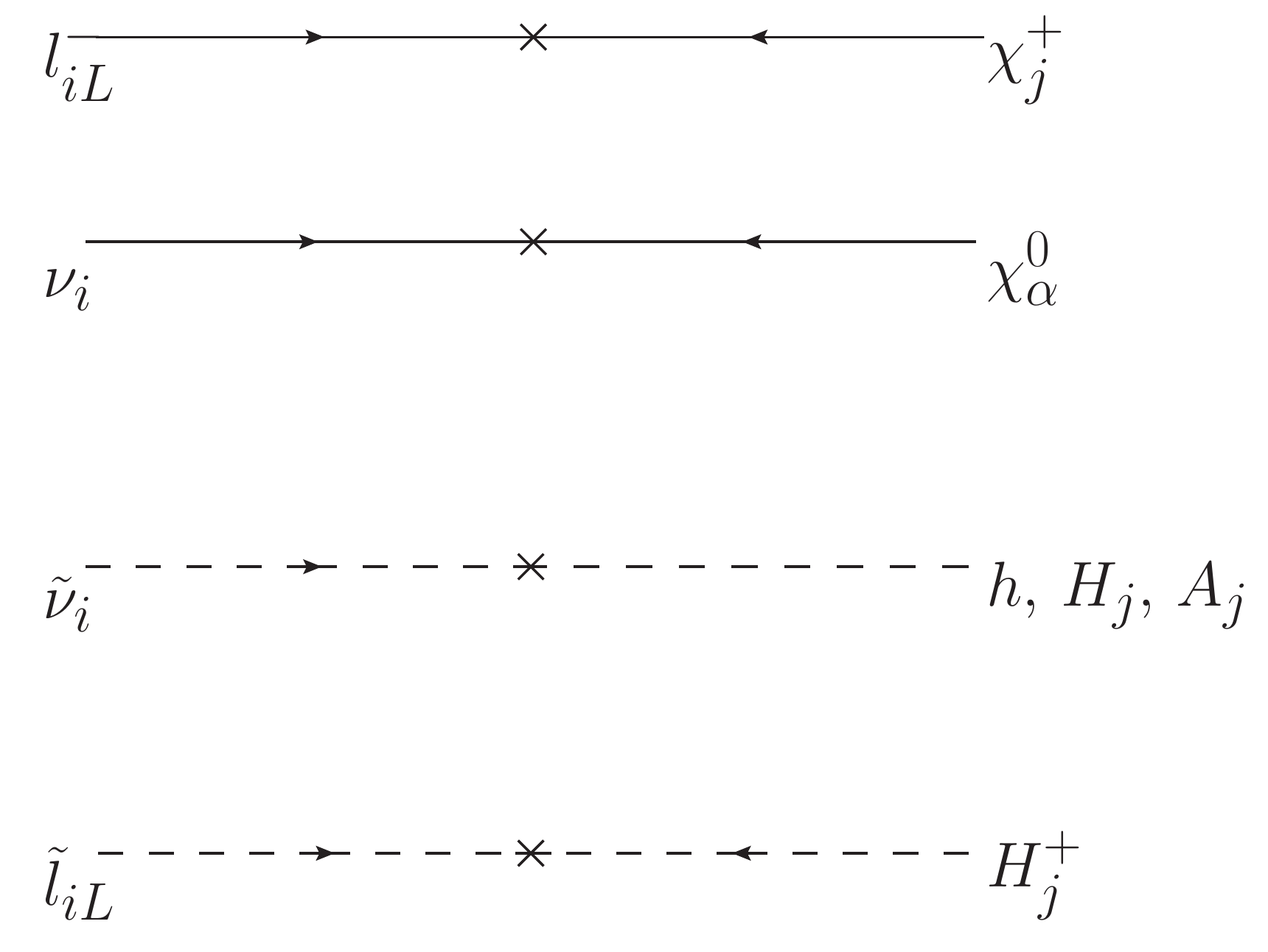}
%\caption{ Bilinear RPV vertices. }
%\label{RPVbilinears}
%\end{figure}
  \end{minipage}
  \begin{minipage}{0.45\linewidth}
   \vspace{-10pt} \bea
&&i\tilde \mu_{ij}^+\equiv i(\mu_{1i} Z_+^{2j}+\mu_{2i}Z_+^{3j})\label{tildemu}
\\\nonumber\\
&&i\tilde \mu_{i\alpha}\equiv i(\mu_{1i} Z_N^{2\alpha}+\mu_{2i}Z_N^{3\alpha})\\\nonumber\\
&&\frac{i}{\sqrt 2}\tilde B_{i\{h,H_j,A_j\}}\equiv \frac{i}{\sqrt 2}\left[B_{1i}\{Z_R^{00},Z_R^{0j}, i Z_H^{0j}\}\right.\nonumber\\
&&+B_{2i}\{Z_R^{10},Z_R^{1j},i Z_H^{1j}\}+(M_{\tilde L}^2)_{0(1+i)}\{Z_R^{20},Z_R^{2j},i Z_H^{2j}\}\label{BiH}\nonumber\\
&&+\left.(M_{\tilde L}^2)_{1(1+i)}\{Z_R^{30},Z_R^{3j},iZ_H^{3j}\}\right]\\\nonumber\\
&&i\tilde B_{ij}\equiv i\left(B_{1i}Z_H^{0j}+B_{2i}Z_H^{1j}+(M_{\tilde L}^2)_{0(1+i)}Z_H^{2j}\right.\nonumber\\
&&+\left.(M_{\tilde L}^2)_{1(1+i)}Z_H^{3j}\right)\label{Bik}
\eea    
  \end{minipage}

\vspace*{3mm}
\noindent where we used
\beq
(M_{\tilde L})_{im}=\lambda_{0(1+i)m}\frac{v_{d_1}}{\sqrt 2}+\lambda_{1(1+i)m}\frac{v_{d_2}}{\sqrt 2}.\label{ml}
\eeq

\newpage

The trilinear RPV vertices which include two Higgs fields arise from
\eq{newterm} and are represented below
\vspace*{3mm}

\begin{minipage}{0.45\linewidth}   
%    \begin{figure}[H]
\includegraphics[scale=.80]{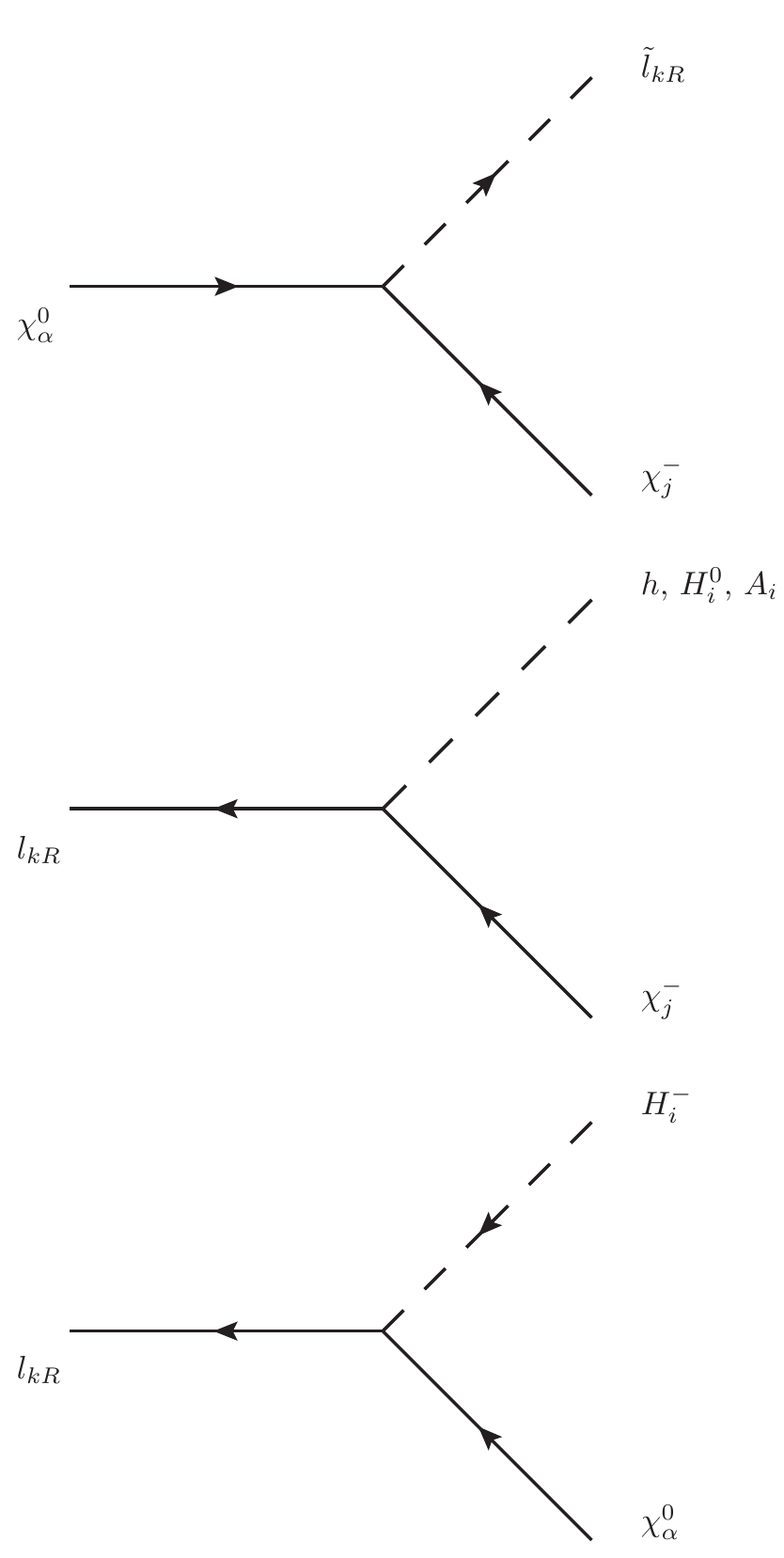}
%\caption{ Trilinear RPV vertices including two Higgs fields. }
%\label{2Higgs RPV vertices}
%\end{figure}
  \end{minipage}
  \begin{minipage}{0.45\linewidth}
   \vspace{-10pt} \bea
&&i \tilde\lambda_k (Z_-^{3j}Z_N^{4\alpha}-Z_-^{2j}Z_N^{5\alpha})\\\nn\\\nn\\\nn\\\nn\\\nn\\\nn\\
&&\frac{i}{\sqrt 2}\tilde\lambda_k (Z_-^{3j}\{Z_R^{20},Z_R^{2i},iZ_H^{2i}\}-Z_-^{2j}\{Z_R^{30},Z_R^{3i},iZ_H^{3i}\})\nn\\\\\nn\\\nn\\\nn\\\nn\\\nn\\
&& i\tilde\lambda_k (Z_H^{3i}Z_N^{4\alpha}-Z_H^{2i}Z_N^{5\alpha})
\eea    
  \end{minipage}

\vspace*{3mm}
\noindent  
  where $\tilde \lambda_k=\lambda_{01k}=-\lambda_{10k}$.

The triliniear R-parity conserving vertices involving a neutrino are
\vspace*{3mm}

\begin{minipage}{0.45\linewidth}   
%\begin{figure}[H]
\includegraphics[scale=0.8]{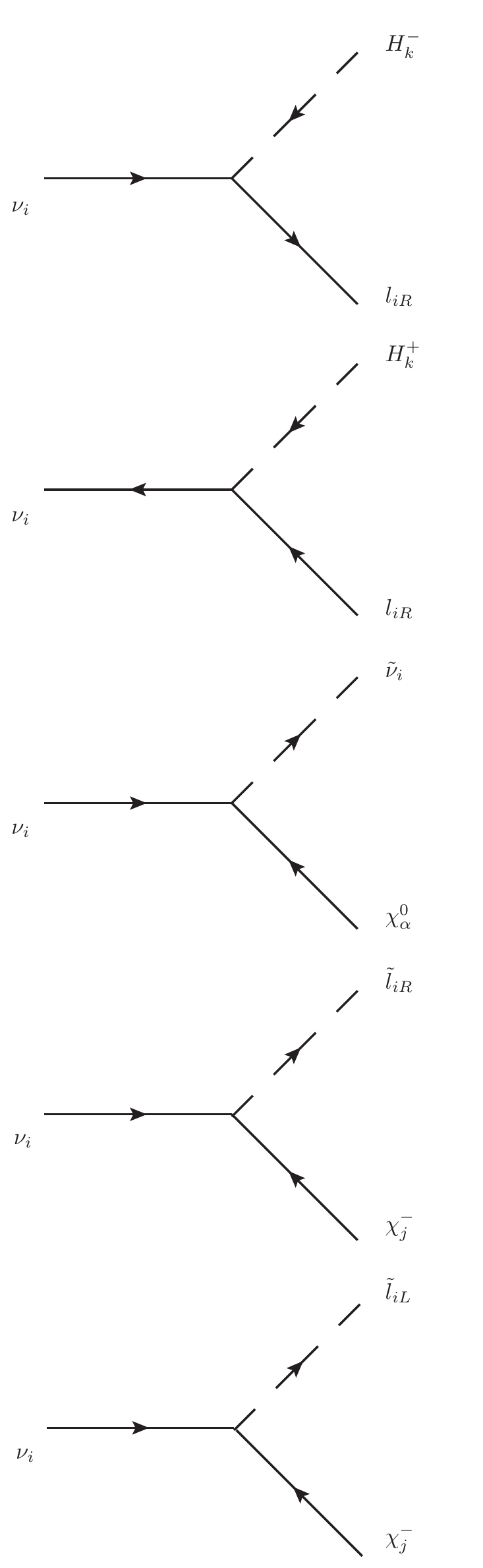}
%\caption{Feynman rules for the R-parity conserving trilinear vertices involving one neutrino.}
%\label{neutrinovertices}
%\end{figure}
 \end{minipage}
  \begin{minipage}{0.45\linewidth}
  \vspace{-40pt}
  \bea
  &&i  (\lambda_{0(1+i)i}Z_H^{1k}+\lambda_{1(1+i)i}Z_H^{2k})\\\nn\\\nn\\\nn\\\nn\\\nn\\
  &&i  (\lambda_{0(1+i)i}Z_H^{1k}+\lambda_{1(1+i)i}Z_H^{2k})\\\nn\\\nn\\\nn\\\nn\\\nn\\
  &&-i\frac{g}{ 2}\left(Z_N^{1\alpha}-\frac{g'}{g}Z_N^{0\alpha}\right)\\\nn\\\nn\\\nn\\\nn\\\nn\\
  &&  i(\lambda_{0(1+i)i}Z_-^{2j}+\lambda_{1(1+i)i}Z_-^{3j})\\\nn\\\nn\\\nn\\\nn\\\nn\\
  &&-igZ_-^{1j}
  \eea
  \end{minipage}

\vspace*{3mm}
\newpage
There are other R-parity conserving vertices which we have also extended to include them in our diagrams. They are :\\

\bigskip

\begin{minipage}{0.35\linewidth}   
%\begin{figure}[H]
\includegraphics[scale=0.75]{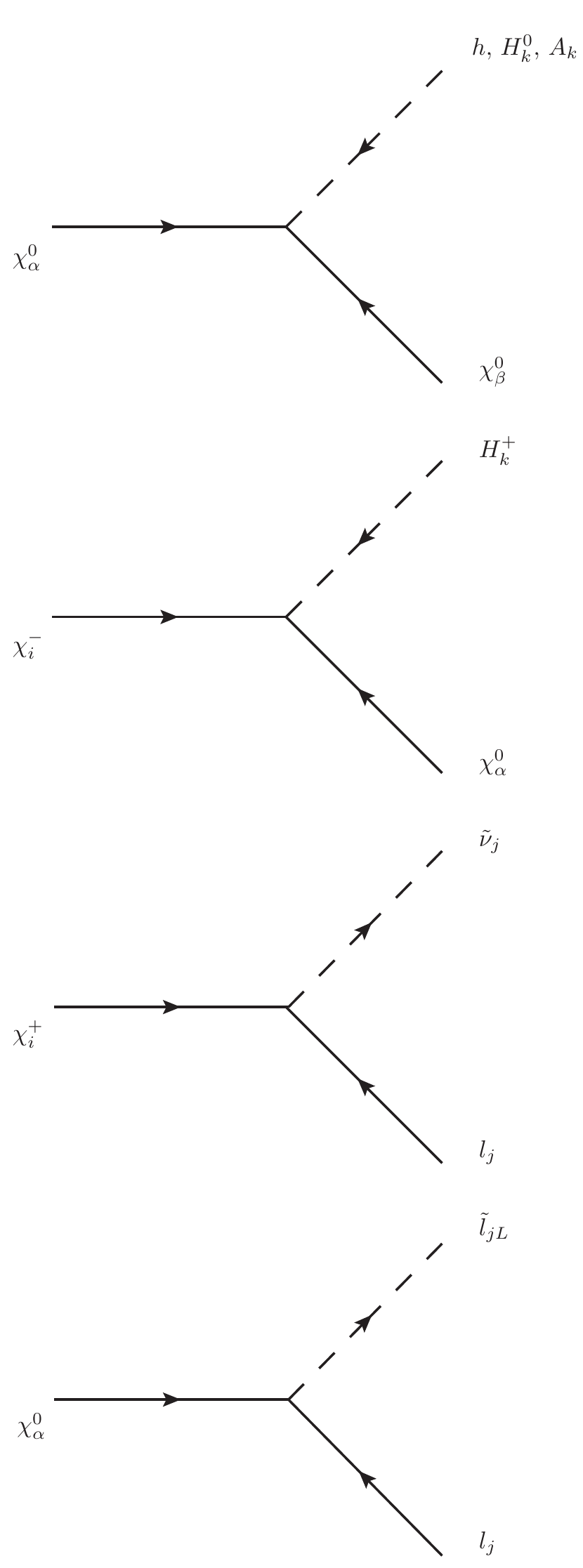}
%\caption{Other trilinear vertices}
%\label{others}
%\end{figure}
 \end{minipage}
  \begin{minipage}{0.45\linewidth}
  \vspace{-30pt}
  \bea
  &&-i\frac{g}{2\sqrt 2}\left[\left(\{Z_R^{20},Z_R^{2k},iZ_H^{2k}\}Z_N^{4\alpha}
+\{Z_R^{30},Z_R^{3k},iZ_H^{3k}\}Z_N^{5\alpha}
\right.\right.\nn\\
&&\left.\left.-\{Z_R^{00},Z_R^{0k},iZ_H^{0k}\}Z_N^{2\alpha}-\{Z_R^{10},Z_R^{1k},iZ_H^{1k}\}Z_N^{3\alpha}\right)\right.\nn\\
&&\left.\left(Z_N^{1\beta}-Z_N^{0\beta}\frac{g'}{g}\right)+\left(\{Z_R^{20},Z_R^{2k},iZ_H^{2k}\}Z_N^{4\beta} \right.\right.\nn\\
&&\left.\left.+\{Z_R^{30},Z_R^{3k},iZ_H^{3k}\}Z_N^{5\beta}-\{Z_R^{00},Z_R^{0k},iZ_H^{0k}\}Z_N^{2\beta}\right.\right.\nn\\ &&\left.\left.-\{Z_R^{10},Z_R^{1k},iZ_H^{1k}\}Z_N^{3\beta}\right)\left(Z_N^{1\alpha}-Z_N^{0\alpha}\frac{g'}{g}\right)\right]\nn\\\\\nn\\
 &&-ig(Z_H^{2k}+Z_H^{3k})\left[(Z_N^{4\alpha}+Z_N^{5\alpha})Z_-^{1i}\right.\nn\\
 &&\left.-\frac{1}{\sqrt 2}\left(Z_N^{1\alpha}+\frac{g'}{g}Z_N^{0\alpha}\right)(Z_-^{2i}+Z_-^{3i})\right]\\\nn\\\nn\\\nn\\\nn\\\nn\\\nn\\
 &&-igZ_+^{1i}\\\nn\\\nn\\\nn\\\nn\\\nn\\\nn\\
 &&-i\frac{g}{\sqrt 2}\left(Z_N^{1\alpha}-\frac{g'}{g}Z_N^{0\alpha}\right)
  \eea
  \end{minipage}

\newpage
%%%%%%%%%%%%%%%%%%%%%%%%%%%%%%%%%%%%%%%%%%%%%%%%%%%%%%%%%%%%%%%%%%%%%%%%%%%%%%%%%%%%%%
\section{One loop contributions to neutrino masses}\label{oneloop}

Here we collect the results for the one-loop diagrams that contribute
to the neutrino masses but do not include the new term in \eq{newterm}
that we have studied in this work. Due to the extra Higgs fields, the
results are not exactly what we have in the MSSM, and thus we show them
here.

\begin{figure}[t]
     \begin{center}
        \subfigure[$\lambda\lambda$ loop]{%
            \label{ll}
            \includegraphics[width=0.47\textwidth]{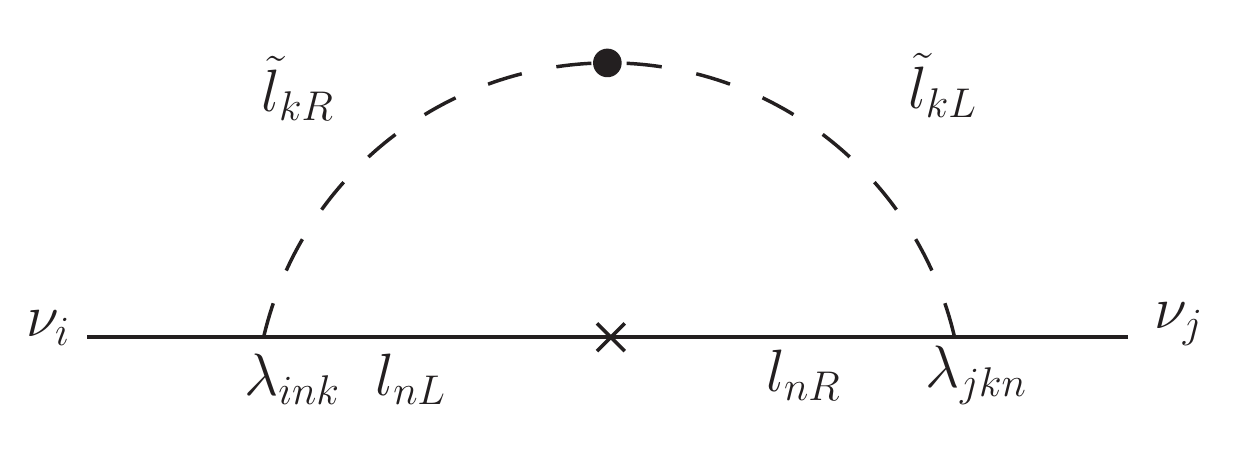} }
        \subfigure[$\lambda'\lambda'$ loop]{%
           \label{llq}
           \includegraphics[width=0.47\textwidth]{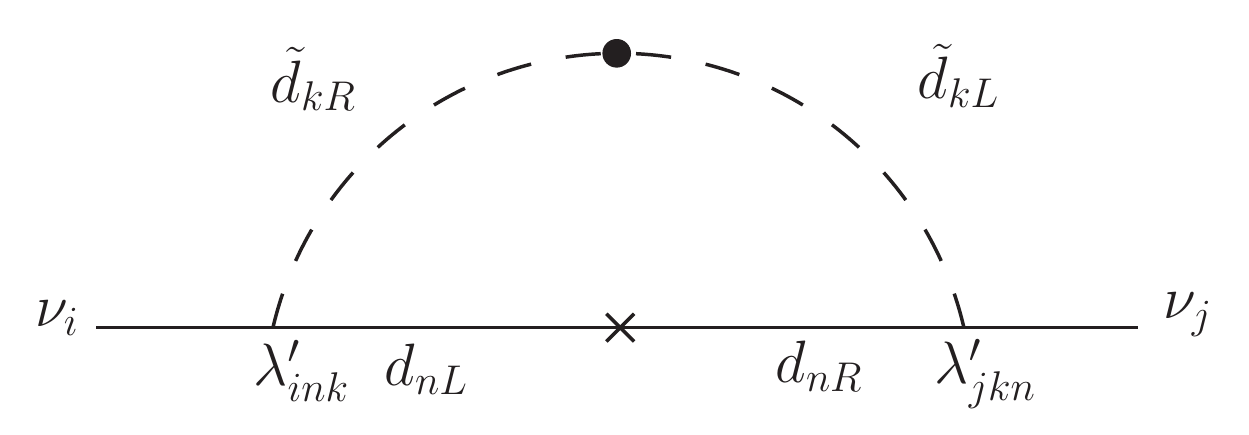} }
    \end{center}
    \caption{%
        $\lambda\lambda$ and $\lambda'\lambda'$ loops.
     }%
   \label{fig:ll}
\end{figure}
The contributions coming from trilinear RPV couplings, which have been already studied in the literature are represented in \fig{fig:ll}. Approximate expressions for them, which are enough for our study are:

\beq
\delta {m_\nu}_{ij}^{\lambda\lambda}\approx\frac{1}{8\pi^2}\sum_{n,k}\lambda_{ink}\lambda_{jkn}\frac{m_{l_n}\Delta m^2_{\tilde l_k}}{m^2_{\tilde l_k}}
\eeq

\beq
\delta {m_\nu}_{ij}^{\lambda'\lambda'}\simeq\frac{3}{8\pi^2}\sum_{n,k}\lambda'_{ink}\lambda'_{jkn}\frac{m_{d_n}\Delta m^2_{\tilde d_k}}{m^2_{\tilde d_k}}
\eeq

\begin{figure}[b]
     \begin{center}
        \subfigure[BB contribution to the neutrino mass matrix]{%
            \label{bb loops}
            \includegraphics[width=0.47\textwidth]{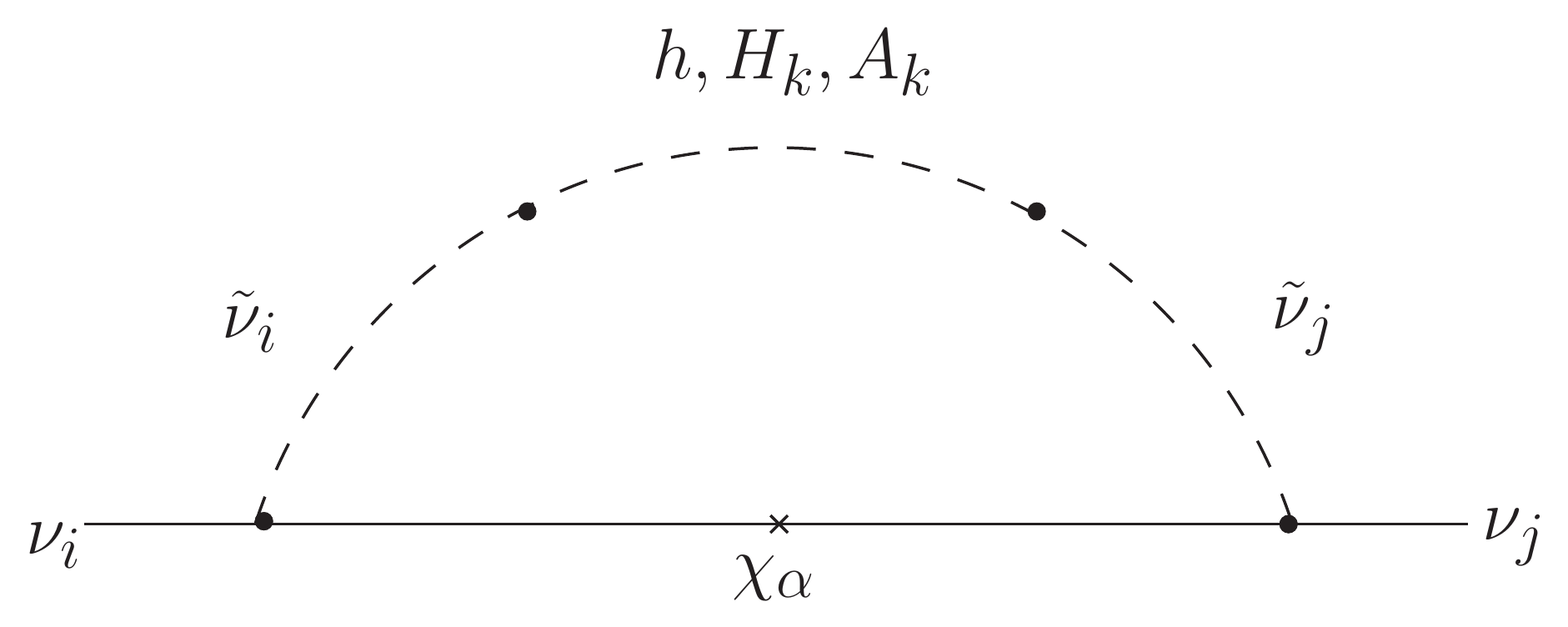} }
        \subfigure[$\mu$B contribution to the neutrino mass matrix]{%
           \label{muB loops}
           \includegraphics[width=0.47\textwidth]{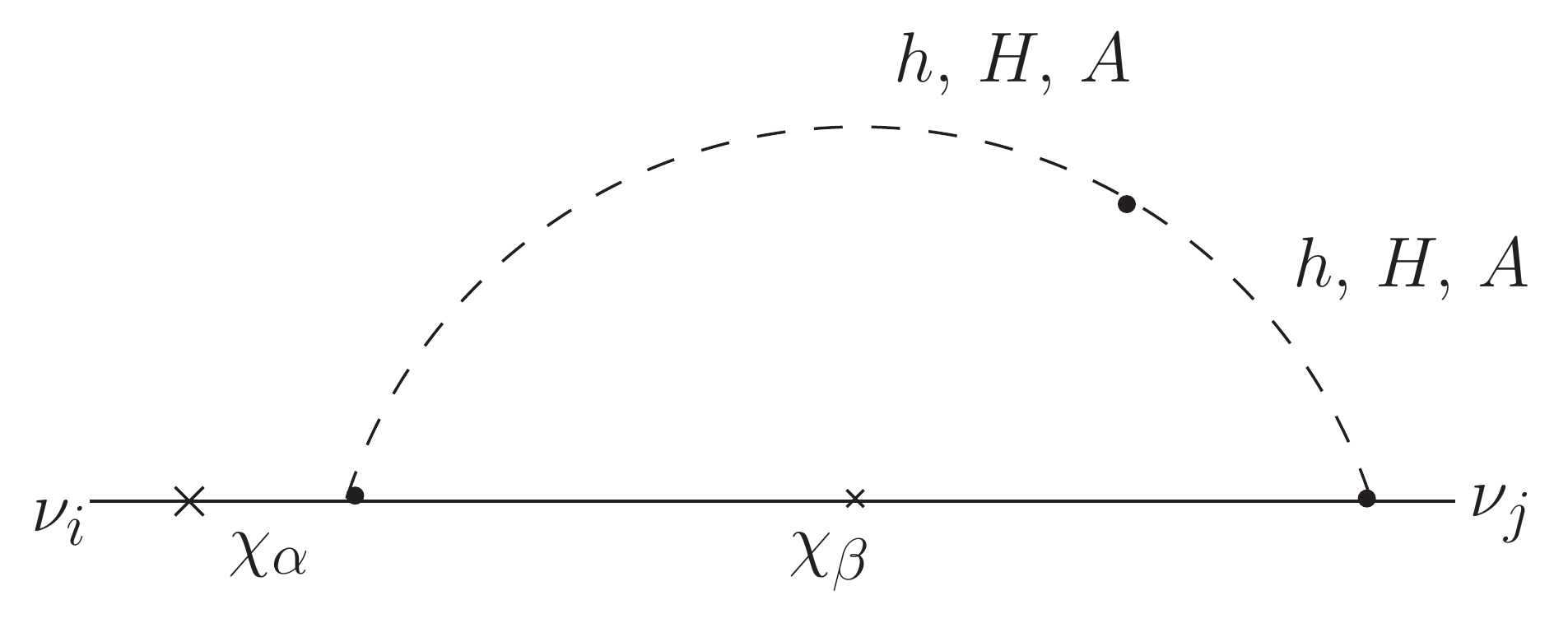} }
    \end{center}
   \caption{%
        $BB$ and $\mu B$ loops.
     }%
   \label{fig:subfigures}
\end{figure}

The soft supersymmetric breaking RPV terms combined in $\tilde B_{ik}$ and $\tilde B_{i\{h,H_j,A_j\}}$, defined in Eqs. (\ref{Bik}) and (\ref{BiH}) respectively, also produce contributions to the neutrino masses at the loop level as represented in \fig{bb loops}.

The one loop contribution is given by
\bea
\delta {m_\nu}_{ij}^{BB}&=&\sum_\alpha \frac{g^2}{4}\left(Z_N^{0\alpha}-\frac{g'}{g}Z_N^{1\alpha}\right)^2\left[\tilde B_{ih}\tilde B_{jh}I_4(m_h,m_{\tilde\nu_i},m_{\tilde\nu_j},m_{\chi_\alpha})\right.\nonumber\\
&+&\left.\sum_k \tilde B_{iH_k}\tilde B_{jH_k}I_4(m_{H_k},m_{\tilde\nu_i},m_{\tilde\nu_j},m_{\chi_\alpha})+\sum_k \tilde B_{iA_k}\tilde B_{j A_k}I_4(m_{A_k},m_{\tilde\nu_i},m_{\tilde\nu_j},m_{\chi_\alpha})\right]
\label{1LBB}\eea
where $I_4$ is defined in \eq{I4}.

Finally, we study the $\tilde\mu \tilde B$ Loops represented in \fig{muB loops}. These kind of loops contribute like:
\bea
\delta {m_\nu}_{ij}^{\mu B}&=&\sum_{\alpha,\beta}\frac{g^2}{4}\tilde\mu_{i\alpha} \left(Z_N^{0\alpha}-\frac{g'}{g}Z_N^{1\alpha}\right)\frac{m_{\chi_\beta}}{m_{\chi_\alpha}}\left\{\tilde B_{jh}\left[\left(Z_R^{20}Z_N^{4\alpha}+Z_R^{30}Z_N^{5\alpha}-Z_R^{10}Z_N^{3\alpha}\right)\frac{}{}\right.\right.\nonumber\\
&&\left.\left.\left(Z_N^{1\beta}-\frac{g'}{g}Z_N^{0\beta}\right)+\left(Z_R^{20}Z_N^{4\beta}+Z_R^{30}Z_N^{5\beta} -Z_R^{00}Z_N^{2\beta}-Z_R^{10}Z_N^{3\beta}\right)\left(Z_N^{1\alpha}-\frac{g'}{g}Z_N^{0\alpha}\right)\right]\right.\nonumber\\
&&\left.I_3(m_{\chi_\beta},m_{\tilde\nu},m_h)+\sum_k \tilde B_{jH_k}\left[\left(Z_R^{2k}Z_N^{4\alpha}+Z_R^{3k}Z_N^{5\alpha}\right)\left(Z_N^{1\beta}-\frac{g'}{g}Z_N^{0\beta}\right)\right.\right.\nonumber\\
&&\left.\left.+\left(Z_R^{2k}Z_N^{4\beta}+Z_R^{3k}Z_N^{5\beta} -Z_R^{0k}Z_N^{2\beta}-Z_R^{1k}Z_N^{3\beta}\right)\left(Z_N^{1\alpha}-\frac{g'}{g}Z_N^{0\alpha}\right)\right]\right.\nonumber\\
&&\left.I_3(m_{\chi_\beta},m_{\tilde\nu},m_{H_k})+\sum_k \tilde B_{jA_k}\left[\left(Z_H^{2k}Z_N^{4\alpha}+Z_H^{3k}Z_N^{5\alpha}\right)\left(Z_N^{1\beta}-\frac{g'}{g}Z_N^{0\beta}\right)\right.\right.\nonumber\\
&&\left.\left.+\left(Z_H^{2k}Z_N^{4\beta}+Z_H^{3k}Z_N^{5\beta} -Z_H^{0k}Z_N^{2\beta}-Z_H^{1k}Z_N^{3\beta}\right)\left(Z_N^{1\alpha}-\frac{g'}{g}Z_N^{0\alpha}\right)\right]\right.\nonumber\\
&&\left.I_3(m_{\chi_\beta},m_{\tilde\nu},m_{A_k})\frac{}{}\right\}+(i\leftrightarrow j)
\eea
where $\tilde\mu_{i\alpha}$ is defined in \eq{tildemu}. Note that in this result we have neglected terms which are proportional to the tree-level masses.

\section{Loop integrals}
Here we collect some loop integrals that we have used throughout this work. For all of the integrals a positive and infinitesimal imaginary part is assumed in the propagators.
\bea
I_2(m_1,m_2)&\equiv&\frac{1}{i}\int\frac{d^4p}{(2\pi)^4}\frac{1}{p^2-m_1^2}\frac{1}{p^2-m_2^2}=-\frac{1}{16\pi^2}\frac{m_1^2}{m_1^2-m_2^2}\ln\frac{m_1^2}{m_2^2}\label{I2}\\
I_3(m_1,m_2,m_3)&\equiv&\frac{1}{i}\int\frac{d^4p}{(2\pi)^4}\frac{1}{p^2-m_1^2}\frac{1}{p^2-m_2^2}\frac{1}{p^2-m_3^2}=\frac{1}{m_1^2-m_2^2}\left[I_2(m_1,m_3)-I_2(m_2,m_3)\right].\label{I3}
\eea
When all masses are equal we get:
\beq
I_3( m, m, m)=\frac{1}{32\pi^2}\frac{1}{ m^2}.\label{I3m}
\eeq
Next we have
\bea
I_4(m_1,m_2,m_3,m_4)&\equiv&\frac{1}{i}\int\frac{d^4p}{(2\pi)^4}\frac{1}{p^2-m_1^2}\frac{1}{p^2-m_2^2}\frac{1}{p^2-m_3^2}\frac{1}{p^2-m_4^2}\nonumber\\
\hspace*{35mm}&=&\frac{1}{m_1^2-m_2^2}\left[I_3(m_1,m_3,m_4)-I_3(m_2,m_3,m_4)\right].\label{I4}
\eea
For the case where the masses are equal
\beq
I_4(m,m,m,m)=\frac{-1}{96\pi^2}\frac{1}{m^4}.
\eeq
Moving on 
\bea
&&I_5(m_1,m_2,m_3,m_4,m_5)=\nn\\[5pt]
&&\frac{1}{m_2 m_4 m_5}\int \frac{d^4q}{(2\pi)^4}
\frac{d^4k}{(2\pi)^4}\frac{m_2 m_3 m_5+m_2 k^2-m_5 q^2+(m_4-m_2+m_5)q\cdot k}{(q^2-m_1^2)^2(q^2-m_2^2)((q-k)^2-m_3^2)^2((q-k)^2-m_4^2)(k^2-m_5^2)}\nn\\[5pt]
&&=\frac{\partial}{\partial m_1^2}\frac{\partial}{\partial m_3^2}\left\{\frac{1}{m_1^2-m_2^2}\frac{1}{m_3^2-m_4^2}J_3(m_5, m_2, m_4,m_1,m_3,m_5)\right.\nn\\
&&\left.+\frac{1}{m_1^2-m_2^2}\frac{1}{m_4^2-m_3^2}J_3(m_5, m_2, m_4,m_1,m_4,m_5)+\frac{1}{m_2^2-m_1^2}\frac{1}{m_3^2-m_4^2}J_3(m_5, m_2, m_4,m_2,m_3,m_5)\right.\nn\\
&&\left.+\frac{1}{m_2^2-m_1^2}\frac{1}{m_4^2-m_3^2}J_3(m_5, m_2, m_4,m_2,m_4,m_5)\right\}\,, \label{eq:I5}
\eea
where we have defined
\bea
J_3(m_1,m_2,m_3,m_A,m_B,m_C)&\equiv&\nn\\[5pt]
&&\hspace{-132pt}\frac{1}{m_2 m_4 m_5}\int \frac{d^4q}{(2\pi)^4}
\frac{d^4k}{(2\pi)^4}\frac{m_1 m_2 m_3+m_2 k^2-m_1 q^2+(m_3+m_1-m_2)q\cdot k}{(q^2-m_A^2)((k-q)^2-m_B^2)(k^2-m_C^2)}\nn\\
&&\hspace{-145pt}=\frac{(m_1+m_2+m_3)}{2 m_1 m_2 m_3}T_2(m_A,m_B)-\frac{(m_1-m_2+m_3)}{2 m_1 m_2 m_3}T_2(m_A,m_C)-\frac{(m_1+m_2-m_3)}{2 m_1 m_2 m_3}T_2(m_B,m_C)\nn\\
&&\hspace{-145pt}+\left(1+\frac{m_3 \left(m_A^2-m_B^2+m_C^2\right)+m_1 \left(- m_A^2-m_B^2+m_C^2\right)+m_2 \left(-m_A^2+m_B^2+m_C^2\right)}{2 m_1 m_2 m_3}\right) \nn\\
&&\hspace{-140pt}T_3(m_A,m_B,m_C),
\eea
and
\bea
T_2(m_1,m_2)&\equiv&(2\pi)^{-2D}\int \frac{d^Dk\, d^D
  l}{(k^2-m_1^2)(l^2-m_2^2)}=-\frac{(\mu^2)^{4-D}}{(4\pi)^{D}}\left(\Gamma\left( 1-D/2\right)\right)^2(m_1^2 m_2^2)^{D/2-1}\,\\
T_3(m_1,m_2,m_3)&\equiv&\frac{(\mu^2)^{4-D}}{(2\pi)^{2D}}\int \frac{d^Dk\, d^D
  l}{(k^2-m_1^2)((k-l)^2-m_2^2)(l^2-m_3^2)}\,,\label{T3}
\eea
where $\mu$ is the dimensional regularization scale, and $T_3$ has
been evaluated in \cite{Ford:1992pn}. Note that even though both $T_2$
and $T_3$ diverge and are therefore dimensional regularization scale dependent, the total sum of all their contributions in \eq{eq:I5} is finite and scale independent. The same thing happens for Eqs. (\ref{eq:I6}) and (\ref{eq:I7}).
For the case where all the masses are equal we get
\bea
I_5( m, m, m, m, m, )&=&\frac{1}{(4\pi)^4}\frac{1}{m^6}\left\{-\frac{5}{12}+3\int_0^1 dx(1-x)
x^3\int_0^1 dy\frac{ (1-y)^2 y}{ 2((1-x) x (1-y)+y)^3}\right\}\nn\\
&\approx& -\frac{1}{(4\pi)^4}\frac{0.14}{ m^6}\label{I5m},
\eea
where in the last step the integral is computed numerically.
Next we have
\bea
&&I_6(m_1,m_2,m_3,m_4,m_5,m_6)=\frac{1}{m_2 m_3 m_6}\int \frac{d^4q}{(2\pi)^4}
\frac{d^4k}{(2\pi)^4}\nn\\[5pt]
&&\frac{m_2 m_3 m_6+m_2 k^2-m_6 q^2+(m_3-m_2+m_6)q\cdot k}{(q^2-m_1^2)^2(q^2-m_2^2)((q-k)^2-m_3^2)((q-k)^2-m_4^2)((q-k)^2-m_5^2)(k^2-m_6^2)}\nn\\[5pt]
&&=\frac{\partial}{\partial m_1^2}\left\{\frac{J_3(m_6,m_2,m_3,m_1,m_3,m_6)}{(m_1^2-m_2^2)(m_3^2-m_4^2)(m_3^2-m_5^2)}+\frac{J_3(m_6,m_2,m_3,m_1,m_4,m_6)}{(m_1^2-m_2^2)(m_4^2-m_3^2)(m_4^2-m_5^2)}\right.\nn\\
&&\hspace*{33mm}+\left.\frac{J_3(m_6,m_2,m_3,m_1,m_5,m_6)}{(m_1^2-m_2^2)(m_5^2-m_3^2)(m_5^2-m_4^2)}+\frac{J_3(m_6,m_2,m_3,m_2,m_3,m_6)}{(m_2^2-m_1^2)(m_3^2-m_4^2)(m_3^2-m_5^2)}\right.\nn\\[5pt]
&&\hspace*{33mm}-\left.\frac{J_3(m_6,m_2,m_3,m_2,m_4,m_6)}{(m_2^2-m_1^2)(m_4^2-m_3^2)(m_4^2-m_5^2)}+\frac{J_3(m_6,m_2,m_3,m_2,m_5,m_6)}{(m_2^2-m_1^2)(m_5^2-m_3^2)(m_5^2-m_4^2)}\right\}\,. \label{eq:I6}
\eea
For the case where all the masses are equal we get
\beq
I_6( m, m, m, m, m, m)=I_5(m,m,m,m,m).\label{I6m}
\eeq
Last we have,
\bea
&&I_7(m_1,m_2,m_3,m_4,m_5,m_6,m_7)=\int \frac{d^4q}{(2\pi)^4}
\frac{d^4k}{(2\pi)^4}\nn\\[5pt]
&&
\frac{m_1 m_4 m_7+m_1 k^2-m_7 q^2+(m_4-m_1+m_7)q\cdot k}{(q^2-m_1^2)(q^2-m_2^2)(q^2-m_3^2)((q-k)^2-m_4^2)((q-k)^2-m_5^2)((q-k)^2-m_6^2)(k^2-m_7^2)}\nn\\[5pt]
&&=\frac{J_3(m_7,m_1,m_4,m_1,m_4,m_7)}{(m_1^2-m_2^2)(m_1^2-m_3^2)(m_4^2-m_5^2)(m_4^2-m_6^2)}+\frac{J_3(m_7,m_1,m_4,m_1,m_5,m_7)}{(m_1^2-m_2^2)(m_1^2-m_3^2)(m_5^2-m_4^2)(m_5^2-m_6^2)}\nn\\
&&+\frac{J_3(m_7,m_1,m_4,m_1,m_6,m_7)}{(m_1^2-m_2^2)(m_1^2-m_3^2)(m_6^2-m_4^2)(m_6^2-m_5^2)}+\frac{J_3(m_7,m_1,m_4,m_2,m_4,m_7)}{(m_2^2-m_1^2)(m_2^2-m_3^2)(m_4^2-m_5^2)(m_4^2-m_6^2)}\nn\\
&&+\frac{J_3(m_7,m_1,m_4,m_2,m_5,m_7)}{(m_2^2-m_1^2)(m_2^2-m_3^2)(m_5^2-m_4^2)(m_5^2-m_6^2)}+\frac{J_3(m_7,m_1,m_4,m_2,m_6,m_7)}{(m_2^2-m_1^2)(m_2^2-m_3^2)(m_6^2-m_4^2)(m_6^2-m_5^2)}\nn\\
&&+\frac{J_3(m_7,m_1,m_4,m_3,m_4,m_7)}{(m_3^2-m_1^2)(m_3^2-m_2^2)(m_4^2-m_5^2)(m_4^2-m_6^2)}+\frac{J_3(m_7,m_1,m_4,m_3,m_5,m_7)}{(m_3^2-m_1^2)(m_3^2-m_2^2)(m_5^2-m_4^2)(m_5^2-m_6^2)}\nn\\
&&+\frac{J_3(m_7,m_1,m_4,m_3,m_5,m_7)}{(m_3^2-m_1^2)(m_3^2-m_2^2)(m_6^2-m_4^2)(m_6^2-m_5^2)}\,. \label{eq:I7}
\eea
For the case where all the masses are equal we get
\beq
I_7( m, m, m, m, m, m,m)=I_5(m,m,m,m,m)\label{I7m}.
\eeq

\end{appendices}
%%%%%%%%%%%%%%%%%%%%%%%%%%%%%%%%%%%%%%%%

%%%%%%%%%%%%%%%%%%%%%%%%%%%%%%%%%%%%%%%%

\end{document}